\documentclass[fleqn,twoside]{article}
\usepackage{espcrc2}

\usepackage{graphicx}
\usepackage{epsfig}

\newcommand{\AmS}{{\protect\the\textfont2
  A\kern-.1667em\lower.5ex\hbox{M}\kern-.125emS}}

\title{Neutrino Oscillations with Reactor Neutrinos}

\author{Anatael Cabrera\address
  {Laboratoire Astroparticule et Cosmologie (APC).\\
   10 rue Alice Domont et L\'{e}onie Duquet. 75205 Paris. France.}\thanks{Marie Curie Intra-European Fellow.}
}
\begin{document}

\begin{abstract}
Prospect measurements of neutrino oscillations with reactor neutrinos are reviewed in this document.
The following items are described: neutrinos oscillations status, reactor neutrino experimental strategy, impact of uncertainties on the neutrino oscillation sensitivity and, finally, the experiments in the field.
This is the synthesis of the talk delivered during the NOW2006 conference at Otranto (Italy) during September 2006.
\end{abstract}

\maketitle

\section{Neutrino Oscillations Status}

Almost\footnote{With the exception of the LSND\cite{ref:LSND} experiment.} all experimental evidence~\cite{ref:PDG,ref:Thomas}, ranging many orders of magnitude of parameter space, strongly favours neutrino oscillations to be the dominant mechanism causing neutrino flavour mutations during neutrino propagation.
Neutrino oscillations are the macroscopic manifestation of mixing in the leptonic sector. 
Neutrinos, therefore, interact as weak-force flavour neutrinos ($\nu_{e},\nu_{\mu},\nu_{\tau}$)~\cite{ref:PDG}, while they oscillate during their propagation as mass neutrinos, thought to be at least 3: $\nu_{1}$, $\nu_{2}$ and $\nu_{3}$.
The leptonic mixing is embodied by the corresponding mixing matrix called PMNS matrix causing a non-diagonal free Hamiltonian for neutrinos.
In addition, the massive neutrino spectrum must be non degenerate, since their $\Delta m^{2}$ leads the oscillation phase factor as a function of L/E, as shown in the two $\nu$ formula:

\begin{equation}
P(\nu_{\alpha} \rightarrow \nu_{\beta}) = \sin^{2}{(2 \theta)} \sin^{2}{ \left ( \frac{1.27 \Delta m^{2} L}{E} \right )}
\end{equation}

The PMNS matrix can be parametrised in terms of 3 mixing angles ($\theta_{12},\theta_{13},\theta_{23}$) and a complex CP violating phase ($\delta_{CP}$).
Majorana phases are not observable through neutrino oscillations.
Today, $\theta_{12}$ and $\theta_{23}$ are known to be large (dominating the solar and atmospheric oscillation, respectively), $\theta_{13}$ is known to be small (dominating the effective decoupling between solar and atmospheric oscillations), while $\delta_{CP}$ is unknown.

With the current sensitivity, no evidence for additional mechanisms beyond neutrino oscillations have been yet found~\cite{ref:Comp}.
This possibility will be probed by next generation of high precision experiments characterising the PMNS matrix to unprecedented precision. 
The leptonic CP violation is an important prediction, which could be related to mechanisms contributing to the observed matter/antimatter asymmetry in the Universe.
In order to measure the leptonic CP violation using neutrino oscillations, the mixing angle $\theta_{13}$ needs not to be zero~\cite{ref:White-Paper}.
Although a vanishing $\theta_{13}$ is of interest for flavour physics, it would prevent much of the potential study of the PMNS matrix through neutrino oscillations.

\subsection{Prospect Reactor Neutrinos $\theta_{13}$ Measurements}

\begin{figure}[h]
\begin{center}
\epsfig{file=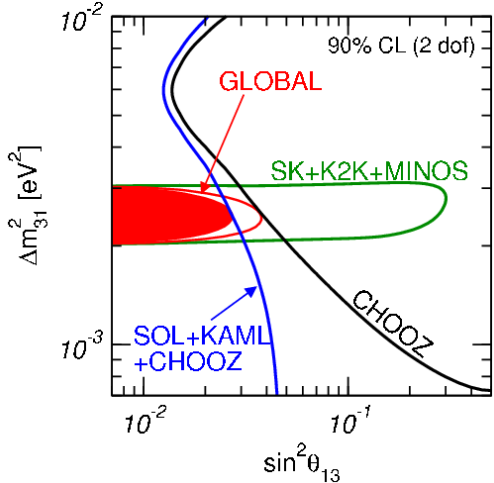,width=0.9\linewidth}
\caption{Limit on $\theta_{13}$ from global analysis~\cite{ref:Thomas}.}
\label{fig:Global}
\end{center}
\end{figure}

Reactor neutrinos have played a critical role to the characterisation of the properties of neutrinos.
This is still true: neutrino reactors could reach the sensitivity necessary to constraint $\sin^{2}(2 \theta_{13}) \sim 0.01$ (10$\times$ better than today~\cite{ref:Thomas}), by a new generation of liquid anti-neutrino detector (LAND)~\cite{ref:White-Paper,ref:Original} experiments: the main subject of this paper.
Our current knowledge on $\theta_{13}$ by global analyses is shown in Figure~\ref{fig:Global}.
The impact of the CHOOZ experiment~\cite{ref:CHOOZ} dominates, although KamLAND~\cite{ref:KamLAND} and solar experiments also contribute.
MINOS~\cite{ref:MINOS} and SuperKamiokande~\cite{ref:SK} contribute mainly through their measurements of $\Delta m^{2}_{atm}$, as this anti-correlates to the limit on $\sin^{2}(2 \theta_{13})$.

\subsection{Neutrino-Beam Complementarity}

Knowledge on $\sin^{2}(2 \theta_{13})$ can be yielded by experiments with $\nu_{\mu}$ beams too~\cite{ref:BeamJapan,ref:BeamEU,ref:BeamUS}.
These experiments are sensitive to unknowns beyond $\theta_{13}$ (unlike reactor experiments), such as $\delta_{CP}$ and, in some cases, the sign of $\Delta m^{2}_{atm}$ since only vacuum oscillations atmospheric data is available.
Therefore, strong complementarity exists between results obtained by both beam and reactor based experiments~\cite{ref:Comp}.
For example, if a reactor neutrino observation was made, the value of $\theta_{13}$ could be fed into the beam neutrino analyses to enhance their sensitivity on $\delta_{CP}$ by reducing the degenerate solution space.
Therefore, it is worthwhile to pursue both approaches with comparable sensitivity for global analyses to infer the most about the structure of the PMNS mixing matrix.

\section{The Reactor Neutrino Scenario}

\subsection{The Inverse-$\beta$ Reaction}

The inverse-$\beta$ reaction ($\bar{\nu_{e}} + p \rightarrow e^{+} + n$) is a critical element of the experimental approach, granting a robust sample of neutrinos (very few analysis cuts) and excellent background rejection.
A neutrino is observed by detecting the correlated prompt-$e^{+}$ and delayed-$n$ energy depositions by a coincidence energy trigger.
The ${\nu}$ energy is accurately inferred from the ${e^{+}}$ energy.
Due to kinematics, there is an energy threshold: $\sim 1.8$~MeV.
This threshold causes the sample of neutrinos detected ($\sim 25\%$ of total) to be those associated to fast decay chains (i.e. high energy $\nu$), which allow the oscillated neutrino sample to be safely isolated from the contribution of slow decay from exhausted-fuel often stored, for some time, near the reactor cores.

\begin{figure}[h]
\begin{center}
\epsfig{file=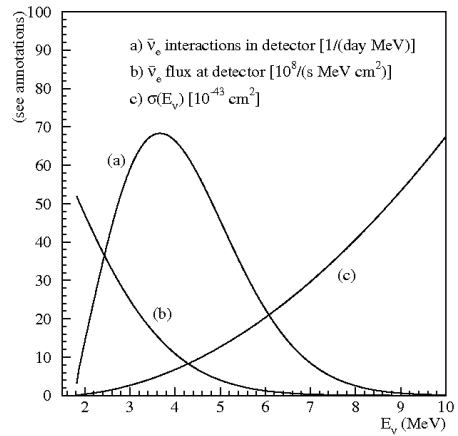,width=0.8\linewidth}
\caption{The measured spectrum by LAND (a) is made of: the reactor neutrino spectrum (b) and inverse-$\beta$ cross-section (c)~\cite{ref:Vogle}} 
\label{fig:Signal}
\end{center}
\end{figure}

Figure~\ref{fig:Signal} shows the LAND observable: the measured reactor neutrino spectrum, which is the convolution of the sum over all the $\beta$-decay spectra of all fission debris and the inverse-$\beta$ cross-section (known within $\sim 0.2\%$)~\cite{ref:Vogle}.
The active volume of this type of detectors consists of liquid scintillator loaded with $0.1\%$ of Gd, since the Gd thermal-neutron capture cross-section is extremely high.
This reduces the time of coincidence cut (about $5\times$), granting further background suppression.
In addition, the neutron-to-Gd capture provides an energy tag: a cascade of gamma-rays amounting to $\sim8$~MeV upon capture - well away from radio-impurities singles. 
Therefore, further background reduction is possible by requiring an energy cut to obtain a neutrino sample through the n-to-Gd capture.

\subsection{Trends of a $\theta_{13}$ Reactor Experiment}

In order to improve our knowledge on $\theta_{13}$, unprecedented high precision is required. 
The statistical and systematic uncertainties should be, at most, $0.5\%$ and $0.6\%$, respectively ($\sim 5\times$ more precise than previous experiments).
Therefore:

\begin{itemize}
\item Near detector makes negligible reactor rate and spectral shape uncertainties.
\item Large (or many) far detector(s) offers larger fiducial volume.
\item Multi-core sites to obtain more $\nu$s.
\item Detector stability: running up to 5 years.
\item Deeper overburden to reduce all cosmogenic backgrounds.
\item Tailored detector design: accurate calibration, detector-to-detector comparison and reduce radioactivity background.
\item Develop robust inter-detector calibration normalisation.
\item Dedicated campaigns to characterise poorly known cosmogenic backgrounds.
\end{itemize}

\subsection{Reactor Flux Uncertainty}

Introducing a near detector allows relative comparison between detectors, hence eliminating the need to know the absolute flux and spectral-shape (Figure~\ref{fig:Signal}) of the detected reactor neutrinos and their time variations due to fuel composition evolution.
The sensitivity to $\sin^{2}(2 \theta_{13})$ of previous reactor experiments was limited by the uncertainty on the rate (and also shape) of the detected neutrino spectrum (known to $\sim 2\%$)~\cite{ref:CHOOZ}.
The oscillation signal will manifest as a deficit on the energy spectrum of the far detector as compared to the normalised near detector spectrum.
The location of the deficit depends on the E/L of the experiment in question, where spectrum maximum is $\sim 4$MeV and L ranges 1-2~km.
Oscillation analyses looking for a difference between overall measured fluxes are said to be {\it rate dominated}, while those aiming for the spectral distortion are said to be {\it shape dominated}.
The latter approach requires much higher bin-to-bin statistics to resolve the dip.
Nonetheless, studies of the reactor neutrino spectrum absolute rate and shape are planned by forthcoming experiments (using their near detectors), as those measurements are important for non-proliferation purposes~\cite{ref:Cribier}.

\begin{figure}[h]
\begin{center}
\epsfig{file=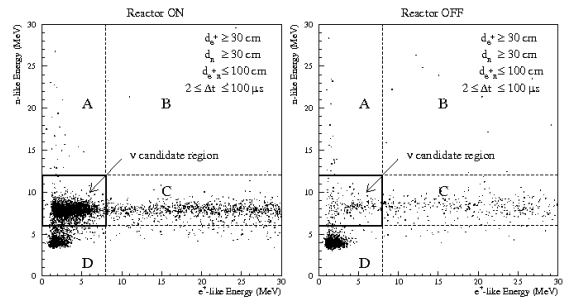,width=1\linewidth}
\caption{Background Characterisation with Reactor on/off Data~\cite{ref:CHOOZ}}
\label{fig:ReactorOFF}
\end{center}	
\end{figure}

\subsection{Backgrounds}

Backgrounds arise from processes mimicking a time coincidence with n-Gd-capture-like emission.
While a radio-pure detector is critical, all dominating backgrounds are associated to cosmic muons, hence, the overburden of each detector relates to its background rate.
Therefore, near and far detectors will have different background (overburden) and signal (distance to cores) rates.
There are 3 types of backgrounds:
{\it i)} {\it accidental} caused by the random coincidence between natural radioactivity ($e^{+}$-like) and the capture of a thermal neutron on Gd ($n$-like).
{\it ii)} {\it correlated} caused by an incoming fast-neutron, which first recoils on a proton ($e^{+}$-like) and, then, gets captured into Gd, once thermalised.
{\it iii)} {\it unstable spallation isotopes} (generated on carbon) cause milliseconds lifetimes $\beta$-$n$ decays, which are impossible to veto.
Figure~\ref{fig:ReactorOFF} shows the spectra of both signal and background (left: reactor-on) and only background (right: reactor-off) for the CHOOZ experiment.
The accidental background is the most dangerous for an oscillation analysis, since its $e^{+}$-like spectrum rises dramatically at low energies, where the oscillation deficit is expected.
Reactor-off data provides a measurement of the effective integrated background spectrum per detector.
This is, however, a very rare occurrence: particularly unlikely in a multi-core reactors.
So, alternative approaches to characterise backgrounds are mandatory.
Background normalisation is generally poorly known, hence dedicated campaigns for their better understanding (at each site) will be carried out.

\subsection{Detector Design}

$\theta_{13}$-LAND has evolved from previous generations, due to the high precision inter-detector comparison envisaged.
The current trend is to use several, multi-volume, cylindrical, small ($<20$t each) detectors - very different, for example, from KamLAND~\cite{ref:KamLAND}.
Much of this trend has been set by the leading experiment in the field: Double Chooz~\cite{ref:DC}.
The RENO~\cite{ref:RENO_NOW06} and Daya Bay~\cite{ref:Daya} collaborations have subscribed to such a trend, while future experiments such as Angra \cite{ref:Angra} may opt for a single large ($500$t) far detector.

Several small detectors meet, at least, one critical goal: cost effective identical near and far detectors.
This is because the near detector need not to be too large due to the high rate of neutrinos granted by the short distance to the reactor core(s).
The far detector limits much of the statistical power of the experiment, so, if high statistics is desired, many small identical far detectors can be used.
Using $N_{F}$ far detectors has the extra advantage that each one can be regarded as an independent experiment, therefore, the overall far detectors uncorrelated uncertainties may scale down with $\sqrt{N_{F}}$.
In addition, Daya Bay has designed their site (tunnels and laboratories) such that detectors can be swapped hoping to better understand detector systematics.

\begin{figure}[h]
\begin{center}
\epsfig{file=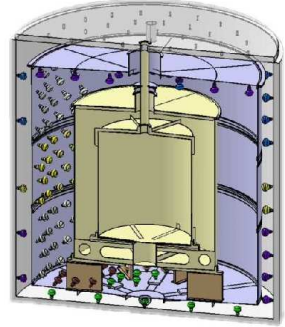,width=0.5\linewidth}
\caption{The Standard Detector (Double Chooz)~\cite{ref:DC}}
\label{fig:Detector}
\end{center}
\end{figure}

Figure~\ref{fig:Detector} shows the 4 volumes standard detector: {\it target} (acrylics Gd loaded liquid scintillator), {\it $\gamma$-catcher} (acrylics unloaded liquid scintillator), {\it buffer} (non-scintillating oil) and {\it inner-veto} (water \v{C}erenkov or liquid scintillator).
There are also an outer inert shield to reject rock radiation and an active tracking {\it outer-veto} for cosmogenic background studies.
The capabilities of this type of detector are:
{\it i)} energy threshold below the $e^{+}$ spectrum ($\sim 0.5$MeV),
{\it ii)} low radioactivity rate (rate in target $<10$Hz),
{\it iii)} detector response uniformity (no position cut and precise energy trigger),
{\it iv)} hardware fiducial volume definition within acrylics (no position cut),
{\it v)} fast-neutron tagging,
{\it vi)} information redundancy (useful for calibration and efficiencies uncertainties).
The same scintillator batches and full readout system ($10"$ or $8"$ PMTs and electronics are used) are used in all detectors: minimise any detector response differences, although most effects are expected to be monitored and corrected by calibration (LED, laser and radioactive sources).


Most importantly, the relative detector normalisation is to be understood to unprecedented accuracy.
3D calibration deployment of the same calibration sources across detectors grants the cancellation of many calibration systematics.
Due to the relative detector comparison, the absolute calibration and normalisation are, in principle, relaxed.

\section{Evolution of the Sensitivity}

Using simulated data, a $\chi^{2}$ function (constructed as the difference of the scaled-near and far detectors data for a certain binning) analysis can be used to study the impact of all possible uncertainties diminishing the sensitivity on $\sin^{2}(2 \theta_{13})$, for $\Delta m^{2}_{atm} = 2.5 \times 10^{-3}$~eV$^{2}$.

\begin{figure}[h]
\begin{center}
\epsfig{file=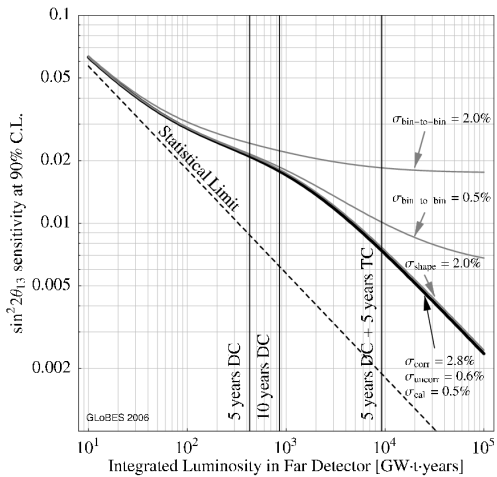,width=0.9\linewidth}
\caption{Evolution of the limit of $\sin^{2}(2 \theta_{13})$~\cite{ref:TC}}
\label{fig:Sevo}
\end{center}
\end{figure}

Figure~\ref{fig:Sevo} summarises the limiting impact of systematics on the limit to $\sin^{2}(2 \theta_{13})$ as a function of the number of $\nu$s in the far detector.
There are 4 domains to be highlighted:
{\it i)} (dashed-curve) the statistical limit improves with $1/ \sqrt{N}$, achieved once the flux uncertainty is eliminated by near detector assuming no systematics.
{\it ii)} (thick-curve) the statistical trend is limited by the inter-detector-normalisation uncertainty, causing a plateau in the sensitivity. 
This trend is caused by the saturation of {\it rate} sensitivity dominating the limit at low statistics ($<10^{3}$GW~t~years).
{\it iii)} (thick-curve) the plateau behaviour turns into a second statistical regime, as the sensitivity becomes {\it shape} dominated, independently of the inter-detector normalisation uncertainty.
The {\it shape} becomes most relevant ($>10^{3}$GW~t~years) as the bin-to-bin statistical power is large enough to resolve distortions caused by a non-vanishing value of $\theta_{13}$. 
{\it iv)} (thinner-curve) once shape uncertainties (i.e. ``bin-to-bin'') are introduced, the sensitivity deviates from second statistical limit.
Shape uncertainties arise from measured spectral differences between the near and far detectors caused by different background contributions or detector systematics.   
In summary, two strategies can be followed to measure $\sin^{2}(2 \theta_{13})$ based on the statistics expected (i.e. number of core(s) and far detector(s)): a rate or a shape dominated measurement.
The knowledge of the inter-detector normalisation (critical for the former) becomes irrelevant for the latter.

\section{Current Reactor $\theta_{13}$ Experiments}

The current collaborations realising a $\theta_{13}$ reactor experiments are Double Chooz~\cite{ref:DC}, RENO~\cite{ref:RENO_NOW06}, Daya Bay~\cite{ref:Daya} and Angra~\cite{ref:Angra} (future) claiming to reach sensitivities on $\sin^{2}(2 \theta_{13})$ up to $0.025$ (by 2012), $0.02$ (by 2013), $0.01$ (by 2013) and $0.006$ (by 2016), respectively.
The KASKA~\cite{ref:KASKA} collaboration has joined Double Chooz, although they still intend a next generation reactor neutrino.
Daya Bay and Angra opt for the high statistics necessary to fully exploit shape information, while the other experiments use both rate and shape. 
RENO is described in~\cite{ref:RENO_NOW06}.
An extension to the Double Chooz experiment is highlighted in~\cite{ref:TC}.
The Hanohano concept was presented in the conference~\cite{ref:HanoNOW,ref:Hano}.

\begin{figure}[h]
\begin{center}
\epsfig{file=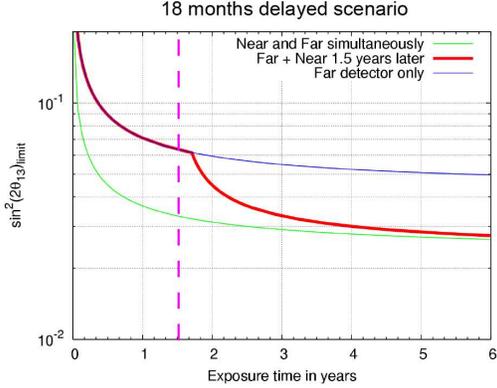,width=0.9\linewidth}
\caption{Double Chooz Sensitivity Evolution~\cite{ref:DCthesis}}
\label{fig:DCevo}
\end{center}
\end{figure}

\subsection{The Double Chooz Experiment}

The Double Chooz collaboration involves France, Germany, Japan, Russia, Spain, UK and US.
Double Chooz is expected to be the first reactor experiment to have a word on $\sin^{2}(2 \theta_{13})$, whose vast R\&D effort is summarised in~\cite{ref:DC}.
The limited size of the far laboratory (former site of CHOOZ) limits the dimensions of the detectors ($8.2$t each, as shown on Figure~\ref{fig:Detector}), while the reuse of this site allows an aggressive time-scale and valuable knowledge about backgrounds from CHOOZ.
Figure~\ref{fig:DCevo} shows the improvement on the limit on $\sin^{2}(2 \theta_{13})$ to $0.06$ (by 2009) and $0.025$ (by 2012), by the far and the near+far, respectively.

\begin{figure}[h]
\begin{center}
\epsfig{file=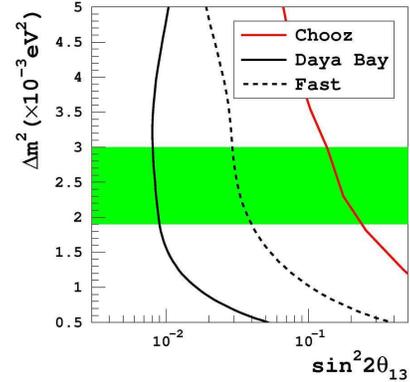,width=0.8\linewidth}
\caption{Daya Bay Sensitivity}
\label{fig:Daya}
\end{center}
\end{figure}

\subsection{The Daya Bay Experiment}

The Daya Bay experiment is made of a collaboration from China, Czech Republic, Russia, and US.
Figure~\ref{fig:Daya} shows the limits that Daya Bay is expected set, also in two phases, due to the time-scale of the civil engineering. 
The first one by 2010 (labelled ``fast'') uses the near ($2\times 20$t) and the middle ($2\times 20$t at $1$km away) detectors, while the second (by 2013) relies on the far ($4\times 20$t between $1.6$ and $1.9$km away) detectors.

\section{Conclusions}

Beyond the promising capability of a measurement (or a limit) of $\sin^{2}(2 \theta_{13})$, many results have already made available through intensive R\&D to control systematic uncertainties to unprecedented levels.
Reactor neutrino experiments are still on the forefront of neutrino research and will provide complementary information to neutrino beams towards the better understanding of the leptonic sector on our Universe.

\end{document}